\documentclass[aps,pra,twocolumn,showpacs,groupedaddress]{revtex4-1}

\begin{document}


\title{Quantum search by partial adiabatic evolution}


\author{Ying-Yu Zhang}
\email[]{newpine2002@yahoo.com.cn}
\author{Song-Feng Lu}
\affiliation{School of Computer Science, Huazhong University of Science and Technology, Wuhan, China.}


\date{\today}

\begin{abstract}
A quantum search algorithm based on the partial adiabatic evolution\cite{Tulsi2009} is provided. We calculate its time complexity by studying the Hamiltonian in a two-dimensional Hilbert space. It is found that the algorithm
improves the time complexity, which is $O(\sqrt{N/M})$, of the local adiabatic search algorithm\cite{Roland2002}, to $O(\sqrt{N}/M)$.
\end{abstract}


\maketitle

\section{Introduction}
Quantum adiabatic computation has attracted a lot of attention in the past decades, such as \cite{Childs2001, Farhi2001, Das2002, Aharonov2004, Ying2007, Mizel2007, Farhi2009, Tulsi2009, Jorg2010}, since it was proposed by Farhi et al.\cite{Farhi2000}. In \cite{Das2002}, an adiabatic algorithm was proposed to solve the Deutsch-Jozsa problem. The algorithm took an exponential time which provided only a quadratic speed up over the best classical algorithm for the problem. A modified algorithm proposed by Wei and Ying \cite{Ying2006} improved the performance to constant time. In \cite{Aharonov2004}, quantum adiabatic computation was proved to be polynomially equivalent to the quantum circuit model. The proof showed that adiabatic quantum computation using Hamiltonians with long-range five-or three-body interactions, or nearest-neighbor two-body interactions with six-state particles, could efficiently simulate the circuit model. This results was soon modified to qubits with two-body interactions \cite{Siu2005}. A simpler proof of the Equivalence was presented in \cite{Mizel2007}. In \cite{Altshuler2009}, quantum adiabatic computation was applied to solve random instances of NP-complete problems. A research outline of its application to solve NP-complete problems can also be found in the paper.

A typical quantum adiabatic algorithm starts with the ground state of the initial Hamiltonian $H_i$, and evolves slowly to the ground state of the final Hamiltonian $H_f$. The system that implements the algorithm uses the time-dependent Hamiltonian
\begin{equation}
\label{Hs}
H(s(t))=(1-s(t))H_i+s(t)H_f.
\end{equation}
The running time (evolution time) is essentially determined by \cite{Tulsi2009}
\begin{equation}
T\geq \Theta(\frac{1}{g^{2}_{min}}),
\end{equation}
where
\begin{equation}
 g_{min}=min[E(1,s)-E(0,s)], (0\leq s \leq 1).
\end{equation}
$E(0,s)$ and $E(1,s)$ are the two lowest eigen values of $H(s)$.
Roland and Cerf \cite{Roland2002} considered the unstructured search problem \cite{Grover1997}, and designed a quantum search algorithm based on local adiabatic evolution. In the algorithm, $H_i$ and $H_f$ are specified as
\begin{equation}
\label{Hi}
H_i=1-|\Psi\rangle\langle\Psi|,
\end{equation}
and
\begin{equation}
\label{Hf}
H_f=1-|\beta\rangle\langle\beta|,
\end{equation}
where
\begin{equation}
\label{Psi}
|\Psi\rangle=\frac{1}{\sqrt{N}}\sum_{i=0}^{N-1}|i\rangle,
\end{equation}
and
$|\beta\rangle$ is an equal superposition of all marked states.
The algorithm can find one marked item in a running time of order $\sqrt{N/M} $.

In \cite{Tulsi2009}, Tulsi proposed a partial adiabatic evolution with $H_f$ a one-dimensional projector Hamiltonian. It was also checked in the paper that Roland and Cerf's results (of the case that there is only one marked state) can be obtained as a special case of the partial adiabatic evolution. In this paper, we give a specified quantum search algorithm based on the partial adiabatic evolution, and show that the algorithm provides a better time complexity, which is $O(\sqrt{N}/M)$, than that, which is $O(\sqrt{N/M})$, of the local adiabatic search algorithm.

This paper is organized as follows. In section 2, we specify the partial adiabatic search algorithm. In section 3, we calculate its time complexity. We conclude the paper in section 4.

\section{Partial adiabatic search algorithm}
In this section, we specify the partial adiabatic evolution \cite{Tulsi2009} as a search algorithm. For convenience, we use different notations. The search algorithm executes the four steps below:

(1) The initial state is prepared to be $|\Psi\rangle$.

(2) At t=0, the Hamiltonian is suddenly changed to $H(s^{-})$ without disturbing the state $|\Psi\rangle$.

(3) The Hamiltonian evolves from $H(s^{-})$ to $H(s^{+})$ linearly in time over duration $T'$.

(4) Measure the state of the system.

Repeat these four steps until we find a marked state.

For the algorithm, $s^{-}$ and $s^{+}$ are specified as $s^{-}=\frac{1}{2}-\frac{1}{2\sqrt{N}}$ and $s^{+}=\frac{1}{2}+\frac{1}{2\sqrt{N}}$. Because the algorithm evolves adiabatically only within a small time interval $[s^{-},s^{+}]$, it is called a partial adiabatic search algorithm.  After step 2, the system that implements the algorithm will be still in the state $|\Psi\rangle$. That is, the system state will be the ground state of $H(s^{-})$ with probability $|\langle\Psi|E(0,s^{-})\rangle|^{2}$. The adiabatic theorem \cite{Messiah1999} guarantees that it will be the ground state of $H(s^{+})$ with probability $|\langle\Psi|E(0,s^{-})\rangle|^{2}$ after step 3. Measuring the state of the system will give the ground state of $H_f=1-|\beta\rangle\langle\beta|$ with probability $P=|\langle\Psi|E(0,s^{-})\rangle|^{2}\times|\langle\beta|E(0,s^{+})\rangle|^{2}$. We call P one-round success probability, and accordingly $T'$ one-round evolution time. The overall time complexity(evolution time) of the algorithm is $T=T'/P$. Here, $T'=\omega/g^{2}_{min}$, where $\omega=s^{+}-s^{-}=\frac{1}{\sqrt{N}}$.

\section{Time complexity}
\subsection{minimum energy gap}
Let
\begin{eqnarray}
|\alpha\rangle=\frac{1}{\sqrt{N-M}}\sum_{x\not\in S}|x\rangle,\\
|\beta\rangle=\frac{1}{\sqrt{M}}\sum_{x\in S}|x\rangle,
\end{eqnarray}
where S is the set of the marked states, and M is the number of the marked states. Throughout of this paper, we suppose S is not empty. This means that $1\leq M\leq N$. The state $|\Psi\rangle$ can be rewritten as \cite{Nielsen2000}
\begin{equation}
|\Psi\rangle=\sqrt{\frac{N-M}{N}}|\alpha\rangle+\sqrt{\frac{M}{N}}|\beta\rangle.
\end{equation}
It is easily to check that any state orthogonal to $|\alpha\rangle$ and $|\beta\rangle$ is a eigen state of $H(s)$ and the corresponding eigen value is 1 which is (N-2) times degenerated. This means that $E(k,s)(k=0,1)$ and are in the subspace spanned by $|\alpha\rangle$ and $|\beta\rangle$. Besides, all the following calculations involve states only in the two-dimensional subspaces panned by $|\alpha\rangle$ and $|\beta\rangle$. So, we can work in the subspace, instead of working in the N-dimensional Hilbert space.

Let the eigen spectrum of H(s) be
\begin{equation}
\label{eigenspectrum}
H(s)|E(k,s)\rangle=E(k,s)|E(k,s)\rangle,
\end{equation}
where $E(k,s)$ and $|E(k,s)\rangle$ are the k-level eigen value and eigen state of $H(s)$ respectively. Throughout this paper, we only consider the two lowest eigenstates and eigenvalues, i.e. k=0,1.
Left multiplying $\langle\alpha|$ to Eq.(\ref{eigenspectrum}), we get
\begin{equation}
\label{leftAlphaToEigenSpectrum}
\langle\alpha|H(s)|E(k,s)\rangle=E(k,s)\langle\alpha|E(k,s)\rangle.
\end{equation}
Substituting the Eq.(\ref{Hs}) into the left side of Eq.(\ref{leftAlphaToEigenSpectrum}), we also get
\begin{eqnarray}
\label{substHsToEigenSpectrum}
\langle\alpha|H(s)|E(k,s)\rangle=\nonumber\\
\langle\alpha|E(k,s)\rangle-(1-s)\langle\alpha|\Psi\rangle\langle\Psi|E(k,s)\rangle,
\end{eqnarray}
with $\langle\alpha|\beta\rangle=0$.
Combining Eqs.(\ref{leftAlphaToEigenSpectrum}) with (\ref{substHsToEigenSpectrum}), we obtain
\begin{equation}
\label{alphaDotProductE}
\langle\alpha|E(k,s)\rangle=\frac{(1-s)\langle\alpha|\Psi\rangle\langle\Psi|E(k,s)\rangle}{1-E(k,s)},
\end{equation}
when $1-E(k,s)\neq 0$.
Similarly,
\begin{equation}
\label{betaDotProductE}
\langle\beta|E(k,s)\rangle=\frac{(1-s)\langle\beta|\Psi\rangle\langle\Psi|E(k,s)\rangle}{1-s-E(k,s)},
\end{equation}
when $1-s-E(k,s)\neq 0$.
Thus, with Eqs.(\ref{alphaDotProductE}) and (\ref{betaDotProductE}), we get
\begin{eqnarray}
\label{psiDotProductE}
\langle\Psi|E(k,s)\rangle=\langle\Psi|\alpha\rangle\langle\alpha|E(k,s)\rangle+\langle\Psi|\beta\rangle\langle\beta|E(k,s)\rangle \nonumber \\
=(\frac{(1-s)A}{1-E(k,s)}+\frac{(1-s)B}{1-s-E(k,s)})\langle\Psi|E(k,s)\rangle,
\end{eqnarray}
where $A=|\langle\Psi|\alpha\rangle|^{2}=\frac{N-M}{N}$ and $B=|\langle\Psi|\beta\rangle|^{2}=\frac{M}{N}$. Rearranging Eq.(\ref{psiDotProductE}), we finally obtain the secular function for H(s)
\begin{equation}
E^{2}(k,s)-E(k,s)+s(1-s)A=0.
\end{equation}
So,
\begin{equation}
E(0,s),E(1,s)=\frac{1\pm\sqrt{1-4s(1-s)A}}{2},
\end{equation}
and
\begin{equation}
g_{s}=E(1,s)-E(0,s)=\sqrt{1-4s(1-s)A}.
\end{equation}
The minimum energy gap $g_{min}=\sqrt{1-A}=\sqrt{M/N}$ is obtained for $s=1/2$. The one-round evolution time is $T^{'}=\sqrt{N}/M$.

\subsection{one-round success probability}
In this subsection, we calculate the one-round success probability. Substituting Eqs.(\ref{alphaDotProductE}) and (\ref{betaDotProductE}) into $|\langle\alpha|E(k,s)\rangle|^{2}+|\langle\beta|E(k,s)\rangle|^{2}=1 (k=0,1)$, we have
\begin{equation}
(\frac{(1-s)^{2}A}{(1-E(k,s))^{2}}+\frac{(1-s)^{2}B}{(1-s-E(k,s))^{2}})|\langle\Psi|E(k,s)\rangle|^{2}=1.
\end{equation}
when $1-E(k,s)\neq 0$ and $1-s-E(k,s)\neq 0$.
If $1-s\neq 0$, this immediately gives
\begin{equation}
\label{psiDotProductE_2}
|\langle\Psi|E(k,s)\rangle|^{2}=\frac{1}{(1-s)^{2}(\frac{A}{(1-E(k,s))^{2}}+\frac{B}{(1-s-E(k,s))^{2}})}.
\end{equation}
Substituting Eq.(\ref{psiDotProductE_2}) into Eq.(\ref{betaDotProductE}), we get
\begin{equation}
\label{betaDotProductE_2}
|\langle\beta|E(k,s)\rangle|^{2}=\frac{B}{(\frac{(1-s-E(k,s))}{1-E(k,s)})^{2}A+B}.
\end{equation}
Because
\begin{eqnarray}
(1-s^{-})^{2}=\frac{1}{4}(1+\frac{1}{\sqrt{N}})^{2}<1,\\
\frac{A}{(1-E(0,s^{-}))^{2}}=\frac{4(N-M)/N}{(1+\sqrt{\frac{1+M}{N}-\frac{M}{N^2}})^{2}}<4,
\end{eqnarray}
and
\begin{eqnarray}
\frac{B}{(1-s^{-}-E(0,s^{-}))^{2}}=\frac{4M}{(1+\sqrt{1+M-\frac{M}{N}})^{2}}\\
\leq\frac{4M}{(1+\sqrt{M})^{2}}<4,
\end{eqnarray}
we obtain
\begin{equation}
|\langle\Psi|E(0,s^{-})\rangle|^{2}>\frac{1}{8}.
\end{equation}
using Eq.(\ref{psiDotProductE_2}).

Because
\begin{eqnarray}
\frac{1-s^{+}-E(0,s^{+})}{1-E(0,s^{+})}=\frac{\frac{-1}{\sqrt{N}}+\sqrt{\frac{M+1}{N}-\frac{M}{N^2}}}{1+\sqrt{\frac{M+1}{N}-\frac{M}{N^2}}}\nonumber\\
<\frac{-1}{\sqrt{N}}+\sqrt{\frac{M+1}{N}-\frac{M}{N^2}}<\sqrt{\frac{2M}{N}}=\sqrt{2B},
\end{eqnarray}
we obtain
\begin{eqnarray}
|\langle\beta|E(0,s^{+})\rangle|^{2}>\frac{B}{2BA+B}=\frac{1}{2A+1}>\frac{1}{3},
\end{eqnarray}
using Eq.(\ref{betaDotProductE_2}).

This provides a lower bound of the one-round success probability
\begin{equation}
P=|\langle\Psi|E(0,\frac{1}{2}-\frac{1}{2\sqrt{N}})\rangle|^{2}\times|\langle\beta|E(0,\frac{1}{2}+\frac{1}{2\sqrt{N}})\rangle|^{2}>\frac{1}{24}.
\end{equation}
So, the overall time complexity of the partial adiabatic search algorithm is $T=T'/P=O(\sqrt{N}/M)$.

\section{Conclusion}
We have provided a quantum search algorithm based on the partial adiabatic evolution. As we have seen that the minimum energy gap along with the one-round success probability determine the overall time complexity of the algorithm. They are calculated by studying the time-dependent Hamiltonian $H(s)$ in a two-dimensional Hilbert space. The overall time complexity is $O(\sqrt{N}/M)$, which provides a speed up of $\sqrt{M}$ over the local adiabatic search algorithm. This is resulted from the fact that the one-round success probability is bounded from below by a constant.

\bibliography{QSPAE}

\begin{thebibliography}{10}%
\makeatletter
\providecommand \@ifxundefined [1]{%
 \ifx #1\undefined \expandafter \@firstoftwo
 \else \expandafter \@secondoftwo
\fi
}%
\providecommand \@ifnum [1]{%
 \ifnum #1\expandafter \@firstoftwo
 \else \expandafter \@secondoftwo
\fi
}%
\providecommand \enquote [1]{``#1''}%
\providecommand \bibnamefont  [1]{#1}%
\providecommand \bibfnamefont [1]{#1}%
\providecommand \citenamefont [1]{#1}%
\providecommand\href[0]{\@sanitize\@href}%
\providecommand\@href[1]{\endgroup\@@startlink{#1}\endgroup\@@href}%
\providecommand\@@href[1]{#1\@@endlink}%
\providecommand \@sanitize [0]{\begingroup\catcode`\&12\catcode`\#12\relax}%
\@ifxundefined \pdfoutput {\@firstoftwo}{%
 \@ifnum{\z@=\pdfoutput}{\@firstoftwo}{\@secondoftwo}%
}{%
 \providecommand\@@startlink[1]{\leavevmode\special{html:<a href="#1">}}%
 \providecommand\@@endlink[0]{\special{html:</a>}}%
}{%
 \providecommand\@@startlink[1]{%
  \leavevmode
  \pdfstartlink
   attr{/Border[0 0 1 ]/H/I/C[0 1 1]}%
   user{/Subtype/Link/A<</Type/Action/S/URI/URI(#1)>>}%
  \relax
 }%
 \providecommand\@@endlink[0]{\pdfendlink}%
}%
\providecommand \url  [0]{\begingroup\@sanitize \@url }%
\providecommand \@url [1]{\endgroup\@href {#1}{\urlprefix}}%
\providecommand \urlprefix [0]{URL }%
\providecommand \Eprint[0]{\href }%
\@ifxundefined \urlstyle {%
  \providecommand \doi [1]{doi:\discretionary{}{}{}#1}%
}{%
  \providecommand \doi [0]{doi:\discretionary{}{}{}\begingroup
  \urlstyle{rm}\Url }%
}%
\providecommand \doibase [0]{http://dx.doi.org/}%
\providecommand \Doi[1]{\href{\doibase#1}}%
\providecommand \bibAnnote [3]{%
  \BibitemShut{#1}%
  \begin{quotation}\noindent
    \textsc{Key:}\ #2\\\textsc{Annotation:}\ #3%
  \end{quotation}%
}%
\providecommand \bibAnnoteFile [2]{%
  \IfFileExists{#2}{\bibAnnote {#1} {#2} {\input{#2}}}{}%
}%
\providecommand \typeout [0]{\immediate \write \m@ne }%
\providecommand \selectlanguage [0]{\@gobble}%
\providecommand \bibinfo [0]{\@secondoftwo}%
\providecommand \bibfield [0]{\@secondoftwo}%
\providecommand \translation [1]{[#1]}%
\providecommand \BibitemOpen[0]{}%
\providecommand \bibitemStop [0]{}%
\providecommand \bibitemNoStop [0]{.\EOS\space}%
\providecommand \EOS [0]{\spacefactor3000\relax}%
\providecommand \BibitemShut [1]{\csname bibitem#1\endcsname}%
\bibitem{Tulsi2009}%
  \BibitemOpen
  \bibfield{author}{%
  \bibinfo {author} {\bibfnamefont{A.}~\bibnamefont{Tulsi}},\ }%
  \bibfield{journal}{%
  \bibinfo {journal} {Physical Review A}\ }%
  \textbf{\bibinfo {volume} {80}},\ \bibinfo {pages} {052328} (\bibinfo {year}
  {2009})%
  \bibAnnoteFile{NoStop}{Tulsi2009}%
\bibitem{Roland2002}%
  \BibitemOpen
  \bibfield{author}{%
  \bibinfo {author} {\bibfnamefont{J.}~\bibnamefont{Roland}}\ and\ \bibinfo
  {author} {\bibfnamefont{N.~J.}\ \bibnamefont{Cerf}},\ }%
  \bibfield{journal}{%
  \bibinfo {journal} {Physical Review A}\ }%
  \textbf{\bibinfo {volume} {65}},\ \bibinfo {pages} {042308} (\bibinfo {year}
  {2002})%
  \bibAnnoteFile{NoStop}{Roland2002}%
\bibitem{Childs2001}%
  \BibitemOpen
  \bibfield{author}{%
  \bibinfo {author} {\bibfnamefont{A.~M.}\ \bibnamefont{Childs}}, \bibinfo
  {author} {\bibfnamefont{E.}~\bibnamefont{Farhi}},\ and\ \bibinfo {author}
  {\bibfnamefont{J.}~\bibnamefont{Preskill}},\ }%
  \bibfield{journal}{%
  \bibinfo {journal} {Physical Review A}\ }%
  \textbf{\bibinfo {volume} {65}},\ \bibinfo {pages} {012322} (\bibinfo {year}
  {2001})%
  \bibAnnoteFile{NoStop}{Childs2001}%
\bibitem{Farhi2001}%
  \BibitemOpen
  \bibfield{author}{%
  \bibinfo {author} {\bibfnamefont{E.}~\bibnamefont{Farhi}}, \bibinfo {author}
  {\bibfnamefont{J.}~\bibnamefont{Goldstone}}, \bibinfo {author}
  {\bibfnamefont{S.}~\bibnamefont{Gutmann}}, \bibinfo {author}
  {\bibfnamefont{J.}~\bibnamefont{Lapan}}, \bibinfo {author}
  {\bibfnamefont{A.}~\bibnamefont{Lundgren}},\ and\ \bibinfo {author}
  {\bibfnamefont{D.}~\bibnamefont{Preda}},\ }%
  \bibfield{journal}{%
  \bibinfo {journal} {Science}\ }%
  \textbf{\bibinfo {volume} {292}},\ \bibinfo {pages} {472} (\bibinfo {year}
  {2001})%
  \bibAnnoteFile{NoStop}{Farhi2001}%
\bibitem{Das2002}%
  \BibitemOpen
  \bibfield{author}{%
  \bibinfo {author} {\bibfnamefont{S.}~\bibnamefont{Das}}, \bibinfo {author}
  {\bibfnamefont{R.}~\bibnamefont{Kobes}},\ and\ \bibinfo {author}
  {\bibfnamefont{G.}~\bibnamefont{Kunstatter}},\ }%
  \bibfield{journal}{%
  \bibinfo {journal} {Physical Review A}\ }%
  \textbf{\bibinfo {volume} {65}},\ \bibinfo {pages} {062310} (\bibinfo {year}
  {2002})%
  \bibAnnoteFile{NoStop}{Das2002}%
\bibitem{Aharonov2004}%
  \BibitemOpen
  \bibfield{author}{%
  \bibinfo {author} {\bibfnamefont{D.}~\bibnamefont{Aharonov}}, \bibinfo
  {author} {\bibfnamefont{W.}~\bibnamefont{van Dam}}, \bibinfo {author}
  {\bibfnamefont{J.}~\bibnamefont{Kempe}}, \bibinfo {author}
  {\bibfnamefont{Z.}~\bibnamefont{Landau}}, \bibinfo {author}
  {\bibfnamefont{S.}~\bibnamefont{Lloyd}},\ and\ \bibinfo {author}
  {\bibfnamefont{O.}~\bibnamefont{Regev}},\ }%
  \bibfield{journal}{%
  \bibinfo {journal} {Proceedings of the 45th Annual IEEE Symposium on
  Foundations of Computer Science(FOCS'04)}}%
   (\bibinfo {year} {2004})%
  \bibAnnoteFile{NoStop}{Aharonov2004}%
\bibitem{Ying2007}%
  \BibitemOpen
  \bibfield{author}{%
  \bibinfo {author} {\bibfnamefont{Z.}~\bibnamefont{Wei}}\ and\ \bibinfo
  {author} {\bibfnamefont{M.}~\bibnamefont{Ying}},\ }%
  \bibfield{journal}{%
  \bibinfo {journal} {Physical Review A}\ }%
  \textbf{\bibinfo {volume} {76}},\ \bibinfo {pages} {024304} (\bibinfo {year}
  {2007})%
  \bibAnnoteFile{NoStop}{Ying2007}%
\bibitem{Mizel2007}%
  \BibitemOpen
  \bibfield{author}{%
  \bibinfo {author} {\bibfnamefont{A.}~\bibnamefont{Mizel}}, \bibinfo {author}
  {\bibfnamefont{D.~A.}\ \bibnamefont{Lidar}},\ and\ \bibinfo {author}
  {\bibfnamefont{M.}~\bibnamefont{Mitchell}},\ }%
  \bibfield{journal}{%
  \bibinfo {journal} {Physical Review Letters}\ }%
  \textbf{\bibinfo {volume} {99}},\ \bibinfo {pages} {070502} (\bibinfo {year}
  {2007})%
  \bibAnnoteFile{NoStop}{Mizel2007}%
\bibitem{Farhi2009}%
  \BibitemOpen
  \bibfield{author}{%
  \bibinfo {author} {\bibfnamefont{E.}~\bibnamefont{Farhi}}, \bibinfo {author}
  {\bibfnamefont{J.}~\bibnamefont{Goldstone}}, \bibinfo {author}
  {\bibfnamefont{D.}~\bibnamefont{Gosset}}, \bibinfo {author}
  {\bibfnamefont{S.}~\bibnamefont{Gutmann}}, \bibinfo {author}
  {\bibfnamefont{H.~B.}\ \bibnamefont{Meyer}},\ and\ \bibinfo {author}
  {\bibfnamefont{P.}~\bibnamefont{Shor}},\ }%
  \bibinfo {journal} {quant-ph/0909.4766v2}%
  \bibAnnoteFile{NoStop}{Farhi2009}%
\bibitem{Jorg2010}%
  \BibitemOpen
\bibfield{journal}{%
    }%
  \bibfield{author}{%
  \bibinfo {author} {\bibfnamefont{T.}~\bibnamefont{J\"{o}rg}}, \bibinfo
  {author} {\bibfnamefont{F.}~\bibnamefont{Krazkala}}, \bibinfo {author}
  {\bibfnamefont{G.}~\bibnamefont{Semerjian}},\ and\ \bibinfo {author}
  {\bibfnamefont{F.}~\bibnamefont{Zamponi}},\ }%
  \bibfield{journal}{%
  \bibinfo {journal} {Physical Review Letters}\ }%
  \textbf{\bibinfo {volume} {104}},\ \bibinfo {pages} {207206} (\bibinfo {year}
  {2010})%
  \bibAnnoteFile{NoStop}{Jorg2010}%
\bibitem{Farhi2000}%
  \BibitemOpen
  \bibfield{author}{%
  \bibinfo {author} {\bibfnamefont{E.}~\bibnamefont{Farhi}}, \bibinfo {author}
  {\bibfnamefont{J.}~\bibnamefont{Goldstone}}, \bibinfo {author}
  {\bibfnamefont{S.}~\bibnamefont{Gutmann}},\ and\ \bibinfo {author}
  {\bibfnamefont{M.}~\bibnamefont{Sipser}},\ }%
  \bibfield{journal}{%
  \bibinfo {journal} {arXiv:quant-ph/0001106v1}}%
   (\bibinfo {year} {2000})%
  \bibAnnoteFile{NoStop}{Farhi2000}%
\bibitem{Ying2006}%
  \BibitemOpen
  \bibfield{author}{%
  \bibinfo {author} {\bibfnamefont{Z.}~\bibnamefont{Wei}}\ and\ \bibinfo
  {author} {\bibfnamefont{M.}~\bibnamefont{Ying}},\ }%
  \bibfield{journal}{%
  \bibinfo {journal} {Physics Letters A}\ }%
  \textbf{\bibinfo {volume} {354}},\ \bibinfo {pages} {271} (\bibinfo {year}
  {2006})%
  \bibAnnoteFile{NoStop}{Ying2006}%
\bibitem{Siu2005}%
  \BibitemOpen
  \bibfield{author}{%
  \bibinfo {author} {\bibfnamefont{S.~A.}\ \bibnamefont{Chin}}\ and\ \bibinfo
  {author} {\bibfnamefont{E.}~\bibnamefont{Krotscheck}},\ }%
  \bibfield{journal}{%
  \bibinfo {journal} {Physical Review A}\ }%
  \textbf{\bibinfo {volume} {71}},\ \bibinfo {pages} {062314} (\bibinfo {year}
  {2005})%
  \bibAnnoteFile{NoStop}{Siu2005}%
\bibitem{Altshuler2009}%
  \BibitemOpen
  \bibfield{author}{%
  \bibinfo {author} {\bibfnamefont{B.}~\bibnamefont{Altshuler}}, \bibinfo
  {author} {\bibfnamefont{H.}~\bibnamefont{Krovi}},\ and\ \bibinfo {author}
  {\bibfnamefont{J.}~\bibnamefont{Roland}},\ }%
  \bibinfo {journal} {quant-ph/0908.2782v2}%
  \bibAnnoteFile{NoStop}{Altshuler2009}%
\bibitem{Grover1997}%
  \BibitemOpen
\bibfield{journal}{%
    }%
  \bibfield{author}{%
  \bibinfo {author} {\bibfnamefont{L.~K.}\ \bibnamefont{Grover}},\ }%
  \bibfield{journal}{%
  \bibinfo {journal} {Physical Review Letters}\ }%
  \textbf{\bibinfo {volume} {79}},\ \bibinfo {pages} {325} (\bibinfo {year}
  {1997})%
  \bibAnnoteFile{NoStop}{Grover1997}%
\bibitem{Messiah1999}%
  \BibitemOpen
  \bibfield{author}{%
  \bibinfo {author} {\bibfnamefont{A.}~\bibnamefont{Messiah}},\ }%
  \emph{\bibinfo {title} {Quantum Mechanics}}\ (\bibinfo {publisher} {Dover},\
  \bibinfo {year} {1999})%
  \bibAnnoteFile{NoStop}{Messiah1999}%
\bibitem{Nielsen2000}%
  \BibitemOpen
  \bibfield{author}{%
  \bibinfo {author} {\bibfnamefont{M.~A.}\ \bibnamefont{Nielsen}}\ and\
  \bibinfo {author} {\bibfnamefont{I.~L.}\ \bibnamefont{Chuang}},\ }%
  \emph{\bibinfo {title} {Quantum Computation and Quantum Information}}\
  (\bibinfo {publisher} {Cambridge University Press},\ \bibinfo {year} {2000})%
  \bibAnnoteFile{NoStop}{Nielsen2000}%
\end{thebibliography}%

\end{document}